\documentclass[11pt]{article}
\usepackage{moriond,epsfig}

\def\be{\begin{equation}}
\def\ee{\end{equation}}
\def\bea{\begin{eqnarray}}
\def\eea{\end{eqnarray}}

\begin{document}
\vspace*{4cm}
\title{CONSTITUENT QUARK and HADRONIC STRUCTURE IN THE NEXT-to-LEADING ORDER
}

\author{ Firooz Arash }

\address{Department of Physics, AmirKabir University, Tafresh Campus, Tehran, Iran 15914\\
Center for Theoretical Physics and Mathematics, AEOI P.O.Box
11365-8486, Tehran, Iran}

\maketitle\abstracts{ The structure of a dressed quark in NLO is
utilized to evaluate the Structure function of Proton and pion. It
is found that there is a simple relationship between $F_{2}^{p}$
and $F_{2}^{\pi}$. The ambiguity in the normalization of
$F_{2}^{\pi}$ is discussed. Furthermore, the polarized structure
function of the constituent(dressed) quark is calculated in the
leading order. While it does produces all the available data on
polarized hadronic structure, it requires a significant
contribution to the proton spin from the orbital angular momentum.
}
The structure function of a Constituent (or dressed) Quark (CQ)
in the Next-to-leading order in QCD is given in\cite{1}. This
structure is assumed to be universal and common to all hadrons.
For a U-type CQ one can write its structure function as:
\begin{eqnarray}
F_{2}^{U}(z,Q^2)=\frac{4}{9}z(G_{\frac{u}{U}}+G_{\frac{\bar{u}}{U}})+
\frac{1}{9}z(G_{\frac{d}{U}}+G_{\frac{\bar{d}}{U}}+G_{\frac{s}{U}}+G_{\frac{\bar{s}}{U}})+...
\end{eqnarray}
where all the functions on the right-hand side are the probability
functions for quarks having momentum fraction $z$ of a U-type CQ
at $Q^{2}$. The moments of these distributions in NLO and at the
initial scales $Q_{0}^{2}=0.283$ $GeV^{2}$ and $\Lambda=0.22$
$GeV$ are calculated in Ref.[1]. The corresponding parton
distributions in a CQ are parameterized as:
\begin{equation}
zq_{\frac{val.}{CQ}}(z,Q^{2})=a z^{b}(1-z)^{c}
\end{equation}
\begin{equation}
zq_{\frac{sea}{CQ}}(z,Q^{2}) = \alpha
z^{\beta}(1-z)^{\gamma}[1+\eta z +\xi z^{0.5}]
\end{equation}
The parameters $a$, $b$, $c$ , $\alpha$, etc. are functions of
$Q^{2}$ and are given in the appendix of Ref.[1]. The structure of
any hadron can be written as:
\begin{equation}
F_{2}^{h}(x,Q^2)=\sum_{CQ}\int_{x}^{1}\frac{dy}{y}
G_{\frac{CQ}{h}}(y)F^{CQ}_{2}(\frac{x}{y},Q^2)
\end{equation}
summation runs over the number of CQ's in a particular hadron.
$F_{2}^{CQ}$ is the CQ structure function whose components are
given in Eqs.(1,2,3). $G_{\frac{CQ}{h}}(y)$ is the probability of
finding a CQ  with the momentum fraction $y$ of the hadron h.
Following Ref.[1] we assume a simple form for the exclusive CQ
distribution in a hadron:
\begin{equation}
G_{Q_{1}...Q_{n}}(y_{1}....y_{n})=R\prod
y_{n}^{a_{n}}\delta(\sum_{n}y_{n}-1)
\end{equation}
where $Q_{n}$ refers to the pertinent constituent (anti)quarks in
a hadron and $y_{n}$ is the corresponding momenta. After
integrating out unwanted momenta, we can arrive at inclusive
distribution of individual CQ. The generic form of these
distributions is as follows.
\begin{equation}
G_{Q/h}(y)=\frac{1}{B(t+1,t+1+u+1)}y^{t}(1-y)^u
\end{equation}
where $B(i,j)$ is Euler Beta functions. For the exact form of
these distributions in a specific hadron see Ref.[1]. In figure
(9) of Ref. [1] we have plotted the $F_{2}^{p}(x, Q^{2})$ for the
range of $Q^{2}=1-2000$ $GeV^{2}$. The predictions
of the model agrees rather well with the experimental data.\\
ZEUS collaboration has also measured the pion structure function
at small $x$ region \cite{2}. the data are normalized in two
different ways, namely; using Pion Flux and additive quark model.
The two normalization differ by a factor of two. ZEUS coll. now
believes that the correct normalization will be somewhere between
the two schemes; being much closer to the additive quark
normalization. I have calculated $F_{2}^{\pi}$ based on CQ
structure in \cite{3} and indeed it yields interesting results. As
an example, in Figure (1), the model results at two values of
$Q^{2}$ are shown. For the entire range of $Q^{2}$ see Ref. [3].
An Interesting observation is made that there is a simple
relationship between $F_{2}^{p}$ and the pion flux normalization
of $F_{2}^{\pi}$(see figure 18 of Ref. [2]); namely
\begin{equation}
F_{2}^{\pi(EF)}(x,Q^2)\approx kF_{2}^{p}(x,Q^2)
\end{equation}
with $k=0.361$. We have calculate the right-hand side of Eq. (7)
in our model and compared it with the effective flux
normalization, $F_{2}^{\pi(EF)}(x,Q^{2})$, in the left hand side.
The comparison is presented in figures (1). As one can see from
the figures, the relationship holds rather well at all $Q^{2}$
values. To avoid any misleading, we emphasize that the direct
calculation of $F_{2}^{\pi}(x,Q^{2})$ in the model (Square points
in figures (1)) is different from $F_{2}^{\pi(EF)}(x,Q^{2})$ and
hence, does not support the effective flux normalization of
$F_{2}^{\pi}$. In other words, our finding merely states that if
we scale $F_{2}^{p}$ by a factor of $k=0.37$
we arrive at equation (7). \\
Since our model produces very good fit to the proton structure
function data in a wide range kinematics, we have also
investigated the relation:
\begin{equation}
F_{2}^{\pi}(x, Q^{2})\approx \frac{2}{3}F_{2}^{p}(\frac{2}{3}x,
Q^{2})
\end{equation}
which is based on color-dipole BFKL-Regee expansion and
corresponds to the ZEUS's additive quark model normalization. Both
sides of the equation (8) is calculated in our model. The results
are shown by the solid lines and square points in Figures (1). It
is worth to note that the following relationship also holds very
well between ZEUS's pion flux normalization data and $F_{2}^{p}$:
\begin{equation}
F_{2}^{\pi(EF)}(x, Q^{2})= \frac{1}{3}F_{2}^{p}(\frac{2}{3}x,
Q^{2})
\end{equation}
which is essentially the same as Eq. (7), except for the factor
$\frac{2}{3}$ in front of $x$ that makes a small correction to the
factor $k$ of Eq. (7). Both Eqs. (7) and (9) indicate that,
regardless of the choice of normalization, the valence structure
of proton, as compared to pion, is shifted to the lower $x$ by a
factor of $\frac{2}{3}$, So that valence $x$ in proton corresponds
to $\frac{2}{3}x$ in pion.
\begin{figure}
\centerline{\begin{tabular}{ccc}
\epsfig{figure=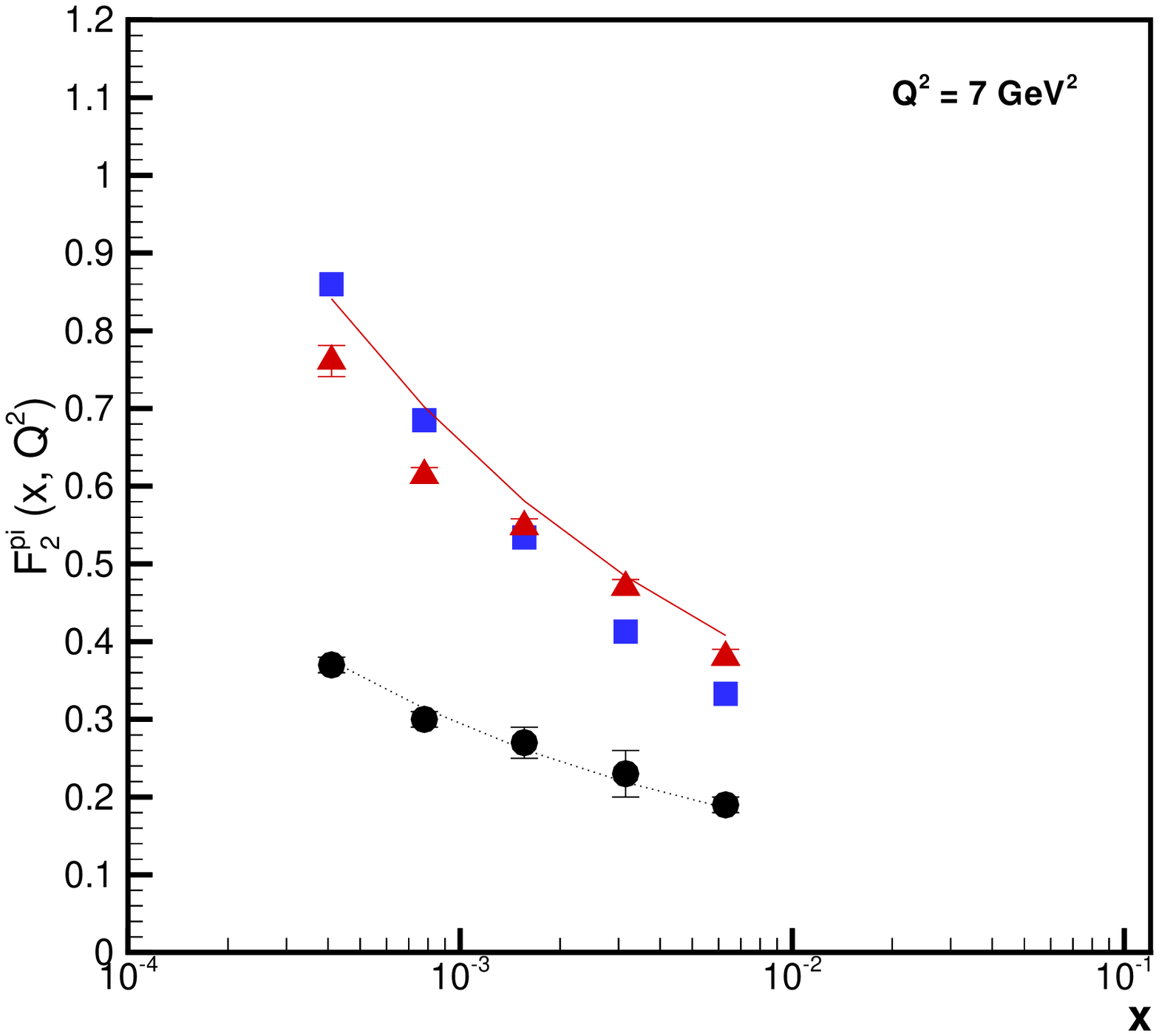,width=5.5cm}
 &\hspace{2cm}&
\epsfig{figure=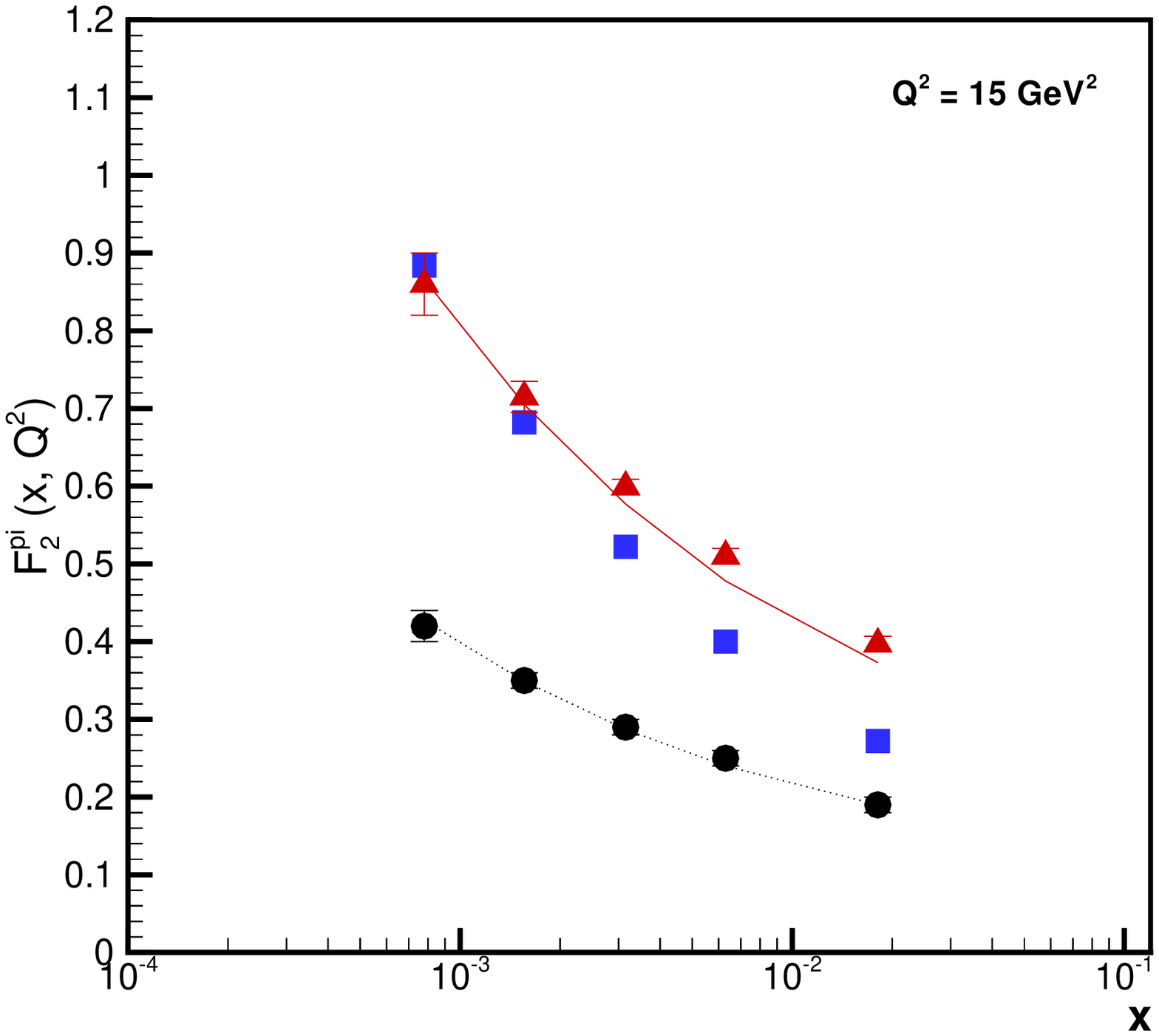,width=5.5cm}
\end{tabular}  }
\caption{\footnotesize $F_{2}^{\pi}$ at $7$ and $15$ $GeV^{2}$.
The circles and triangles are pion flux and additive quark model
normalization of data [1], respectively. Squares are the
calculated results. The solid line represents
$\frac{2}{3}F_{2}^{p}(\frac{2}{3}x, Q^{2})$ and the dotted line is
$0.371 F_{2}^{p}(x, Q^{2})$,both are calculated using the model. }
\label{fig1}
\end{figure}

\section{Polarized Structure function}
Lets denote the polarized structure function of a U-type CQ by
$g_{1}^{U}(z, Q^{2})$ which can be written in terms of its
polarized parton distributions as follows:
\begin{equation}
g_{1}^{U}(z,Q^{2})=\frac{4}{9}(\delta p_{\frac{u}{U}}+ \delta
p_{\frac{\bar{u}}{U}}) +\frac{1}{9}(\delta p_{\frac{d}{U}}+\delta
p_{\frac{\bar{d}}{U}}+\delta p_{\frac{s}{U}}+\delta
p_{\frac{\bar{s}}{U}}) +...
\end{equation}
Denoting the polarized CQ distribution in a hadron by $\Delta
G_{CQ}^{h}(y)$ one can write the polarized structure function of
the hadron as:
\begin{equation}
g_{1}^{h}(x,Q^{2})=\frac{1}{2}\sum_{CQ}\int_{x}^{1} dy \Delta
G_{CQ}^{h}(y) g_{1}^{CQ}(\frac{x}{y},Q^{2})
\end{equation}
The polarized valence quark distribution in a proton at a starting
scale of $Q^{2}_{0}=0.23 GeV^{2}$ is parameterized by GRSV
\cite{4}. We take that parameterization and write
\begin{equation}
\Delta G_{j}(y)=\delta F_{j}(y) G_{j}(y); \hspace{2cm}  j=U, D
\end{equation}
$G_{j}(y)$ are the unpolarized CQ distributions in proton.
Justification for such an ansatz is that at some $Q=Q_{0}$ a
constituent quark behaves as valence quark and its structure
cannot be resolved. Using the results of Ref.[1] the functions
$\delta F_{j}(y)$ is parameterized as follows:
\begin{equation}
\delta F_{j}(y)=
N_{j}y^{\alpha_{j}}(1-y)^{\beta_{j}}(1+\gamma_{j}y+\eta_{j}y^{0.5})
\end{equation}
After evaluating the moments of polarized parton distributions in
and performing an Inverse Mellin Transformation, we arrive at the
polarized parton distribution in a polarized CQ. Thus, For proton
we have:
\begin{equation}
g_{1}^{p}(x,Q^{2})= \frac{1}{2}\int_{x}^{1}dy[2 \Delta
G_{U}(y)g_{1}^{U}(\frac{x}{y},Q^{2}) + \Delta
G_{D}(y)g_{1}^{D}(\frac{x}{y},Q^{2})]
\end{equation}
An interchange of U and D will give the neutron structure
function. In Figure (2) we present the model results along with
the experimental data.
\begin{figure}
\centerline{\begin{tabular}{ccc}
\epsfig{figure=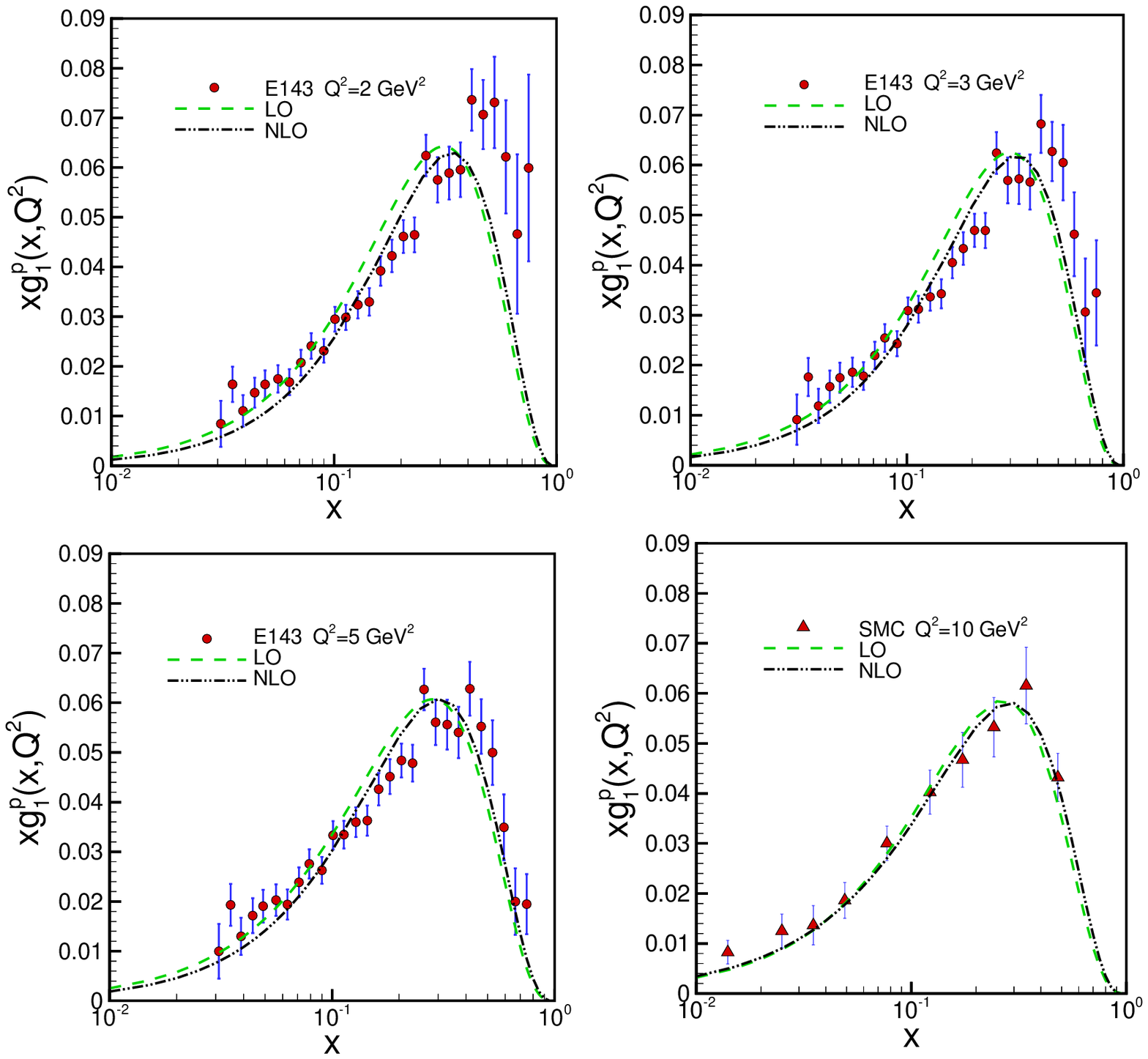,width=5cm}
 &\hspace{2cm}&
\epsfig{figure=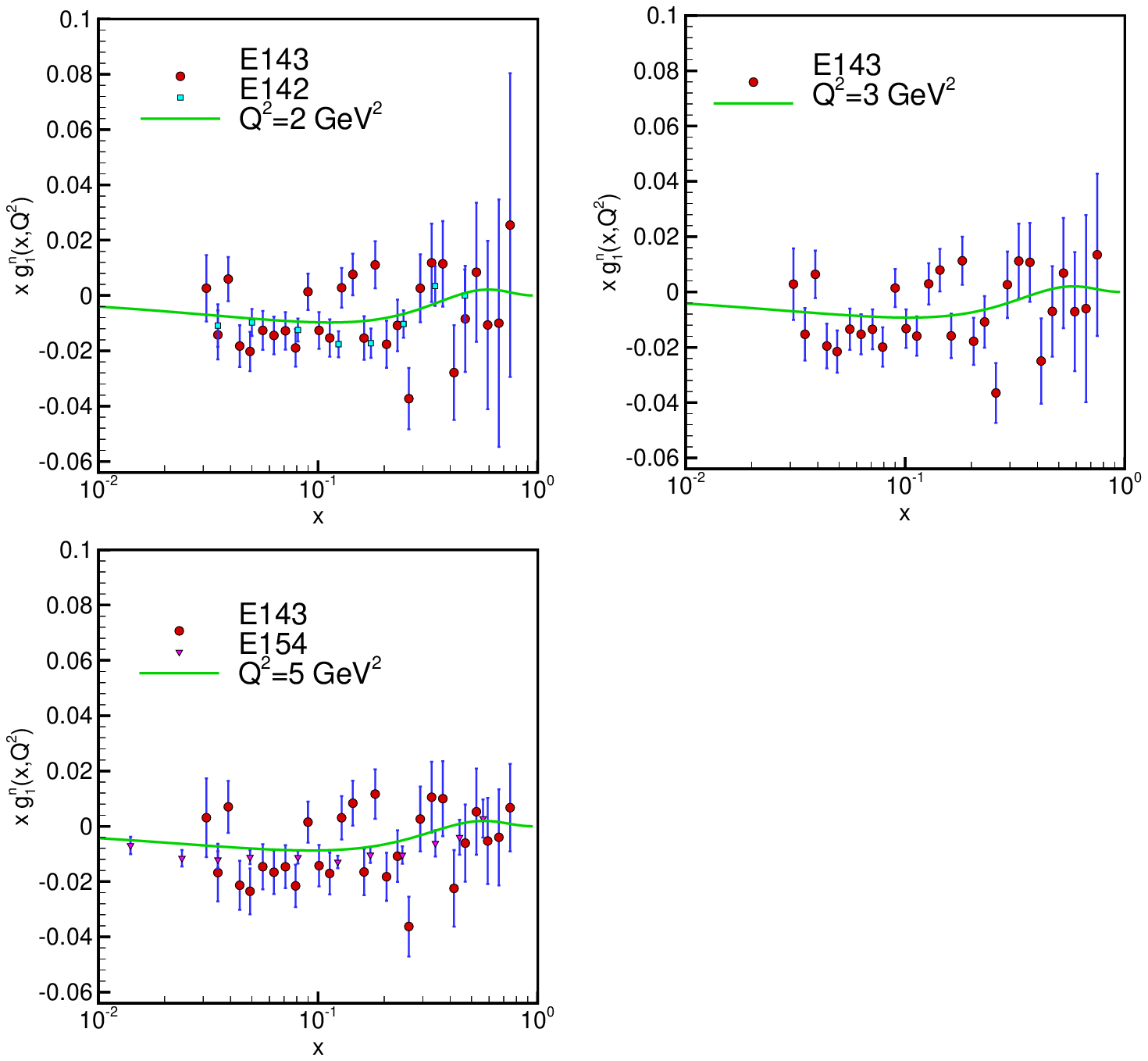,width=5cm}
\epsfig{figure=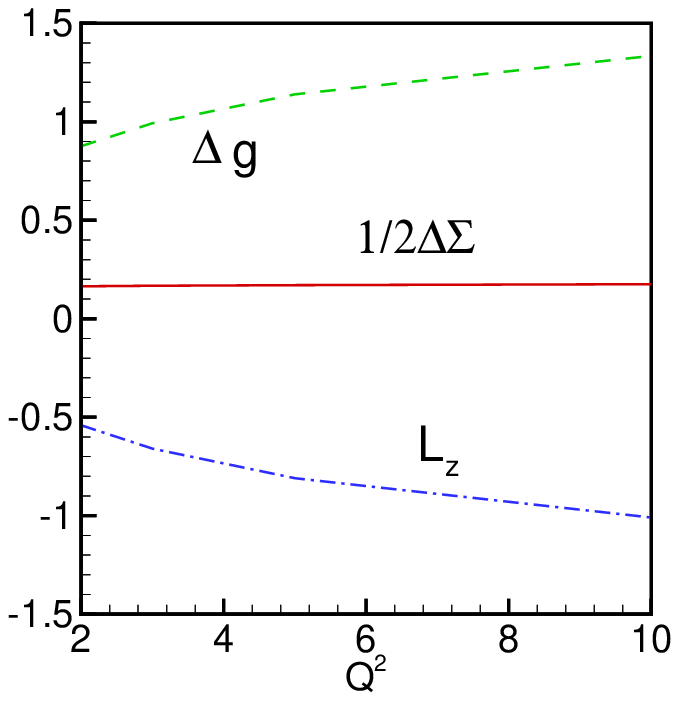,width=5cm}
\end{tabular}  }
\caption{\footnotesize (Left): $g_{1}^{p}(x,Q^{2})$,
(Middle):$g_{1}^{n}(x,Q^{2})$ as calculated from the model and
compared with experimental data; (Right): $L_{z}$ and the gluon
asymmetry in a CQ. } \label{fig2}
\end{figure}
We see that the model easily accounts for the experimental data,
yet the same data does not yield the nucleon spin. Therefore, it
is natural to ask whether the sum of the spins of CQs does produce
the nucleon spin? The answer requires the investigation of the
spin of a CQ itself. Contribution of different components of the
constituent quark to its spin is obtained by the following
integral, say, for a U-type constituent quark:
\begin{equation}
\Delta p_{\frac{i}{U}}(Q^2)=\int^{1}_{0}\delta
p_{\frac{i}{U}}(z,Q^{2}) dz
\end{equation}
where $i$ stands for valence, sea quark and gluon inside a U-type
CQ. For the valence, $\Delta p_{\frac{v}{U}}=1$ for all
$Q^{2}$values, whereas $\Delta p_{\frac{sea}{U}}$ changes with
$Q^{2}$, but remains small around $0.08-0.2$ for a range of
$Q^{2}= 2-10 GeV^{2}$. However $\Delta p_{\frac{gluon}{U}}$ is
fairly large and grows rapidly with the increase of $Q^{2}$
reaching to 4.4 at $Q^{2}= 10 GeV^{2}$. It seems that with these
numbers it is impossible to build a spin $\frac{1}{2}$ constituent
quark just out of quarks and gluon spins. One needs to include the
orbital angular momentum in order to compensate for the growth of
gluon asymmetry. The Orbital angular momentum enters in the
following sum rule
\begin{equation}
S_{z}^{U}=\frac{1}{2}(S_{z}^{v} +S_{z}^{sea})^{U}
+(S^{gluon}_{z})^{U} +L_{z}^{U} =\frac{1}{2}.
\end{equation}
This equation along with the values given for each components
enables us to evaluate $L_{z}$ as depicted in figure (2)

\section{Conclusion}
We have shown that there are basic objects in the hadrons. The
structure of which is calculated in QCD. It provides  good
agreements with experimental data on the structure of hadrons,
both polarized and unpolarized. We have further shown that in
order to account for the spin of nucleon, one needs a large
orbital angular momentum contribution from the partonic structure
of CQ. This contribution is evaluated.

\end{document}